\documentstyle[aps,preprint]{revtex}

\input epsf

\begin{document}

\hfill\vbox{\hbox{\bf CPP-95-11}\hbox{\bf DOE-ER40757-068}
\hbox{\bf NUHEP-TH-95-07}\hbox{July 1995}}\\

\vspace{0.2in}

\begin{center}
{\Large \bf
Strange-Beauty Meson Production at  $p\bar p$ Colliders}\\

\vspace{0.2in}

Kingman Cheung\footnote{Internet address: {\tt cheung@utpapa.ph.utexas.edu}}

{\it Center for Particle Physics, University of Texas, Austin, TX 78712}

Robert J. Oakes\footnote{Internet address: {\tt oakes@fnalv.fnal.gov}}

{\it Department of Physics and Astronomy, Northwestern University, Evanston,
IL 60208}
\end{center}

\begin{abstract}
The production rates and transverse momentum distributions of the
strange-beauty mesons $B_s$ and $B_s^*$ at $p\bar p$ colliders are calculated
assuming fragmentation is the dominant process.  Results are given for the
Tevatron in the large transverse momentum region, where fragmentation is
expected to be most important.
\end{abstract}

\thispagestyle{empty}

\section{Introduction}

Data on the production of strange B mesons is rapidly accumulating.  At LEP
the DELPHI Collaboration has measured the probability for a $\bar b$
antiquark to hadronize into a weakly decaying strange B meson \cite{delphi}.
Assuming the hadronization
is dominated by the fragmentation of the heavy $\bar b$ antiquark,
we determined the
fragmentation functions $D_{\bar b\to B_s}(z,\mu)$ and $D_{\bar b\to B_s^*}
(z,\mu)$ from the measured total probability for a $\bar b$ antiquark to
hadronize into the lowest strange-beauty states
$B_s(^1S_0)$ and $B_s^*( ^3S_1)$ \cite{bs}.
Here $z$ is the fraction of the $\bar b$ antiquark momentum
carried by the $B_s$ or $B_s^*$  at the scale $\mu$.
The momentum distributions of the $B_s$ and $B_s^*$ mesons produced in $e^+e^-$
annihilation at the energy of the $Z$ mass were then predicted \cite{bs}.

Here we extend our calculations to the production of the $B_s$ and $B_s^*$
states in $p\bar p$ annihilation at the Tevatron, using the fragmentation
functions previously determined from the $e^+e^-$ data taken at LEP.
Sec. II presents the cross section for the process in terms of the
parton distribution functions and subprocesses and the fragmentation
functions, which we discuss in Secs. III and IV, respectively.
In Sec. V we present numerical results for the transverse momentum
$(p_T)$ distributions of the strange $B_s$ and $B_s^*$ mesons, as well as the
total cross section for $p_T>p_T^{\rm min}$.
Finally, we discuss these results and summarize our conclusions.

\section{Cross Sections}

In the parton model the cross section for the production by fragmentation
of strange B mesons in proton-antiproton annihilation involves
three main factors: the structure functions of the initial partons in the
proton and antiproton, the subprocess in which the initial partons produce
a particular parton in the final state, and the fragmentation of this final
parton into the strange B meson.  The transverse momentum distribution of the
$B_s$ meson is then of the form
\begin{equation}
\label{bs}
\frac{d\sigma}{dp_T} = \sum_{i,j,k} \int dx_i \int dx_j \int dz \;
f_{i/p}(x_i) f_{j/\bar p}(x_j) \frac{d\hat \sigma}{dp_T} \left(
i(x_i)j(x_j) \to k(\frac{p_T}{z})X ;\; \mu \right) \;
D_{k\to B_s}(z;\mu) \;.
\end{equation}
Here $f_{i/p}(x_i)$ and $f_{j/\bar p}(x_j)$ are the structure functions
of the initial partons $i$ and $j$ carrying fractions of the total
momenta $x_i$ and $x_j$ in the proton and antiproton, respectively.
The production of parton $k$ with momentum $p_T/z$ is described by the
hard subprocess cross section $\hat \sigma$ at the scale $\mu$.  And
$D_{k\to B_s}(z;\mu)$ is the fragmentation function for the parton $k$ to
yield the meson $B_s$ with momentum fraction $z$ at the scale $\mu$.
The factorization scale $\mu$ is of the order of the transverse momentum
of the fragmenting parton.  The soft
physics is contained in the structure functions, while the parton subprocess
is a hard process and can be reliably calculated in perturbation theory.
We shall also assume the fragmentation of $\bar b$ into $B_s$
is also a hard process, an assumption that  will require further discussion
below, since the $s$ quark is not very heavy.

Clearly, there is a formula very similar to Eq.~(\ref{bs}) for the production
of the $B_s^*$, which we need not write explicitly.  Because the radiative
decay $B_s^* \to B_s + \gamma$ is so fast and the $B_s^* - B_s$ mass difference
is so small,  the transverse momentum of $B_s^*$ is hardly affected in
the decay $B_s^* \to B_s +\gamma$, and thus contributes to the inclusive
production of the $B_s$.
We will, therefore, present each individually in the figures
showing $d\sigma/dp_T$ and $\sigma(p_T>p_T^{\rm min})$ and also
include their sum in a table, for convenience.

In the fragmentation process $k\to B_s$ we have included gluons as well as
$\bar b$ antiquarks.  Although the direct $g\to B_s$ fragmentation process does
not occur until order $\alpha_s^3$ there is a significant contribution
coming from the Altarelli-Parisi evolution of the $\bar b$ antiquark
fragmentation
functions from the heavy quark mass to the collider energy scale $Q$,
which is of order $\alpha_s^3 \log(Q/m_b)$ due to the splitting $g\to \bar b$.
This has been shown to be significant for the production of $B_c$ mesons
\cite{induce} and we have also included it here.

\section{Structure Functions and Parton Subprocesses}

In Eq.~(\ref{bs}) we have assumed a factorization scale $\mu$, with the parton
distribution functions being due to soft processes below this scale, while
the parton cross sections for the subprocesses are due to hard contributions
above this scale.
For the parton distribution functions we used the most recent CTEQ
version 3 \cite{cteq}, in which we chose the
leading-order fit, since our calculation is also a leading-order one.
For the subprocesses we used the tree-level cross sections,
to be consistent, since the parton fragmentation functions, which are
discussed in the following section, were
calculated only to leading order.   For the
production of $\bar b$ antiquarks we included the processes
$gg\to b\bar b$, $g\bar b\to g\bar b$, and $q\bar q\to b\bar b$,
while for the production of gluons $g$ we included the processes
$gg\to gg$, $gq(\bar q) \to gq(\bar q)$, and $q\bar q\to gg$, where $q$
denotes any of the quarks, $u,d,s,c,b$.

For the running strong coupling constant $\alpha_s$ we used the simple
one-loop result evolved from its value at $\mu=m_Z$:
\begin{equation}
\alpha_s(\mu) = \frac{\alpha_s(m_Z)}{1+ \frac{33-2n_f}{6\pi} \; \alpha_s(m_Z)
\; \log \left( \frac{\mu}{m_Z} \right ) } \;.
\end{equation}
Here $n_f$ is the number of active flavors at the scale $\mu$ and we chose
$\alpha_s(m_Z)= 0.117$ \cite{pdg}.  From the experimental results on $b$-quark
production cross sections at the Tevatron we know that there are discrepancies
between the calculations and the experiments about a factor of two.
Therefore, adopting a more complicated form for the strong coupling
constant, or using the next-to-leading order subprocess cross sections, or
taking the next-to-leading order structure functions is not critical.
At this point, we are primarily interested in how important the fragmentation
process is for $B_s$ and $B_s^*$ production and how the shapes of the
calculated momentum distributions compare with forthcoming data.

The factorization scale $\mu$ in Eq.~(\ref{bs}) deserves further discussion
before proceeding.
The scale $\mu$ superficially appears to have been invented only to
separate the production of $B_s$ mesons into structure functions,
subprocess cross sections, and fragmentation functions.
In principle, this is so and the production is independent of the choice
of $\mu$.  But in practice, the results do depend on $\mu$, since  only
if all the factors are calculated to all orders in $\alpha_s$ will
the dependence on $\mu$ cancel.
However, for example, the fragmentation functions were calculated only
to leading order and, consequently,  the results will depend on $\mu$.
We shall investigate the sensitivity of the results to the choice of
$\mu$ by varying $\mu$ from $\mu_R/2$ to  $2 \mu_R$, where
$\mu_R=\sqrt{p_T^2({\rm parton}) +m_b^2}$ is our central choice of the scale
$\mu$.
%

\section{Fragmentation Functions}

To obtain the fragmentation functions at the scale $\mu$ we have numerically
integrated the Altarelli-Parisi evolution equations \cite{alt} from the
scale $\mu_0$, which is of the order of the $b$-quark mass.
The evolution equations for the fragmentation functions $D_{\bar b\to B_s}
(z,\mu)$ and $D_{g\to B_s}(z,\mu)$ are
\begin{equation}
\label{Db}
\mu \frac{\partial}{\partial \mu} D_{\bar b\to B_s}(z,\mu) =
\int_z^1 \frac{dy}{y}
P_{\bar b\to \bar b}(z/y,\mu)\; D_{\bar b \to B_s}(y,\mu) +
\int_z^1 \frac{dy}{y} P_{\bar b\to g}(z/y,\mu)\; D_{g \to B_s}(y,\mu) \,,
\end{equation}
and
\begin{equation}
\label{Dg}
\mu \frac{\partial}{\partial \mu} D_{g\to B_s}(z,\mu) = \int_z^1 \frac{dy}{y}
P_{g \to \bar b}(z/y,\mu)\; D_{\bar b \to B_s}(y,\mu) +
\int_z^1 \frac{dy}{y} P_{g \to g}(z/y,\mu)\; D_{g \to B_s}(y,\mu) \,,
\end{equation}
with similar, coupled equations for the $B_s^*$ fragmentation functions.  The
splitting functions $P_{i\to j}(x,\mu)$, to leading order in $\alpha_s(\mu)$,
are explicitly given in Ref.~\cite{induce}.   As boundary conditions in
these calculations we have used the fragmentation functions at the scale
$\mu_0$ calculated  in Ref.~\cite{theory}:
\begin{eqnarray}
D_{\bar b\rightarrow B_s}(z,\mu_0) & = &
\frac{2\alpha_s(2m_s)^2 |R(0)|^2}
{81\pi m_s^3}\; \frac{rz(1-z)^2}{(1-(1-r)z)^6} \nonumber \\
&\times & [ 6 - 18(1-2r)z + (21 -74r+68r^2) z^2  \nonumber \\
 && -2(1-r)(6-19r+18r^2)z^3  + 3(1-r)^2(1-2r+2r^2)z^4 ]\,.
\label{dz1}
\end{eqnarray}
Here $z$ is the fraction of the $\bar b$ antiquark
momentum carried by the $B_s$
meson, $r=m_s/(m_b+m_s)$, and $R(0)$ is the $B_s$ meson $S$-wave radial
wavefunction at the origin.  The corresponding fragmentation function for
$\bar b\to B_s^*$ is \cite{theory}
\begin{eqnarray}
D_{\bar b\rightarrow B_s^*}(z,\mu_0) & = &
\frac{2\alpha_s(2m_s)^2 |R(0)|^2}
{27\pi m_s^3}\; \frac{rz(1-z)^2}{(1-(1-r)z)^6} \nonumber \\
&\times & [ 2 - 2(3-2r)z + 3(3 - 2r+ 4r^2) z^2 \nonumber \\
& &   -2(1-r)(4-r +2r^2)z^3  + (1-r)^2(3-2r+2r^2)z^4 ]\,,
\label{dz2}
\end{eqnarray}

The induced gluon fragmentation functions $D_{g\to B_s,B_s^*}(z,\mu)$ are of
order $\alpha_s^3 \log(\mu/m_b)$ and become important at large values of
the scale $\mu$ relative to the $\bar b$ antiquark fragmentation functions
$D_{\bar b\to B_s,B_s^*}(z,\mu)$, which are of order $\alpha_s^2(\mu)$.
The boundary conditions for the gluon fragmentation functions are
\begin{equation}
D_{g\to B_s}(z,\mu) = D_{g\to B_s^*}(z,\mu)=0
\end{equation}
for $\mu \le 2(m_b+m_s)$, the threshold for producing $B_s$ or $B_s^*$ mesons
{}from a gluon.

The value of $|R(0)|^2$ can be determined from the decay constant $f_{B_s}$
of the $B_s$ meson using the relation \cite{weiss}
\begin{equation}
f_{B_s}^2 = \frac{3}{\pi} \; \frac{|R(0)|^2}{M_{Bs}}
\end{equation}
with $f_{B_s} = 207 \pm 34 \pm 22$ MeV \cite{soni} and $M_{B_s}=5.375$ GeV
\cite{pdg}.

If we choose the $b$ quark mass to be $m_b=5$ GeV the only remaining
parameter in these fragmentation functions is $m_s$.  This strange-quark mass
parameter $m_s$ has been determined previously \cite{bs} using the same
initial fragmentation functions from the experimental value \cite{delphi}
of the probability $f_s^w$ for a $\bar b$ antiquark to fragment into weakly
decaying strange-beauty mesons:
\begin{equation}
f_s^w = \int_0^1 dz \; \left[ D_{\bar b\to B_s}(z,\mu_0) +
                              D_{\bar b\to B_s^*}(z,\mu_0) \right ] \;.
\label{Y}
\end{equation}
Since the total probability for the $\bar b$ antiquark to fragment into
a $B$ meson is independent of scale, the initial scale $\mu_0$ in
Eq.~(\ref{Y}), which is of the order of the $b$-quark mass, was chosen to
be $\mu_0= m_b+2m_s$ as in Ref.~\cite{theory}.
We previously found \cite{bs} $m_s=318$$\scriptsize{\stackrel{\textstyle +47}
{-24}}$ MeV$/c^2$ based on
the DELPHI measurement $f^w_s=0.19\pm 0.06 \pm 0.08$ and assuming
$\alpha_s(m_Z)=0.12$.
Prior to the DELPHI measurement there were older data from LEP \cite{old}
which were consistent with this measurement.
There is now also a new measurement from CDF \cite{cdf} of
the quantity
\begin{equation}
\frac{\sigma(B_s)}{\sigma(B_u) + \sigma(B_d) } = 0.26
\mbox{$\scriptsize{\stackrel{\textstyle +.17}{-.08}}$} \pm .08\;.
\label{X}
\end{equation}
Since the total fragmentation probability $P(b \to {\rm baryon})
\approx 5 - 10$\%, the corresponding value of $f^w_s=P(\bar b \to B_s,B_s^*)$
can be estimated assuming
$P(\bar b \to B_u + B_d + B_s + {\rm baryons})=1$.  Using the newer CDF data,
Eq.~(\ref{X}),  we obtain $f^w_s=0.196 (0.186)$ for $P(\bar b\to {\rm baryon})=
5(10)$\%, which is consistent with the DELPHI measurement.
In the present calculations, therefore, we used the same value
$f_s^w=0.19\pm 0.06 \pm 0.08$ \cite{delphi} but a slightly different value of
$\alpha_s(m_Z)=0.117$ \cite{pdg}.
For the strange quark mass parameter we then obtain
\begin{equation}
m_s = 298 \mbox{$\scriptsize{\stackrel{\textstyle +47}{-23}}$}\; {\rm MeV}
\end{equation}
using the scale-independent relation Eq.~(\ref{Y}).
This value of $m_s$ is not significantly different from the previous
determination and we shall use it in our fragmentation functions
Eqs.~(\ref{dz1}) -- (\ref{dz2}).
The strong coupling constant in Eqs.~(\ref{dz1}) -- (\ref{dz2}) then has
the value  $\alpha_s(2m_s)=0.745$.
While this value of $\alpha_s$ is uncomfortably large  we note that it only
enters, together with $m_s$ and $R(0)$, in the normalization of the
fragmentation functions, which has been determined empirically through
Eq.~(\ref{Y}).

\section{Results and Discussions}

The evolution equations were numerically integrated and the fragmentation
functions at the scale $\mu_R=\sqrt{p_T^2({\rm parton}) + m_b^2}$ were combined
with the structure functions and parton cross sections for the subprocesses
to obtain the cross section, Eq.~(\ref{bs}). Specifically, we calculated
the transverse momentum distributions of $B_s$ and $B_s^*$ mesons produced
in $p\bar p$ collisions at the Tevatron energy $\sqrt{s}=1.8$~TeV.  We
assumed a cut-off on the transverse momentum of $p_T>5$ GeV$/c^2$ and
considered only the rapidity range $|y|<1$.
In Fig.~\ref{fig1} we show the cross section $d\sigma/dp_T$ for both $B_s$ and
$B_s^*$.  To investigate the sensitivity of these results
to the scale $\mu$ we have also
included the results for $\mu=2\mu_R$ and $\mu=\mu_R/2$.
When the scale $\mu$ is less than $\mu_0=m_b+2m_s$, which only happens for the
case of $\mu=\mu_R/2$, we chose the larger of $(\mu,\mu_0)$.
{}From Fig.~\ref{fig1}
it is clear that the choice of scale $\mu$ is not critical for the
transverse momentum distribution; in fact, the variation over the range
$\mu_R/2 <\mu <2\mu_R$ is comparable to the current discrepancies between
the measured and calculated $b$ production cross sections.  As one might
expect, the $(^3S_1) \; B_s^*$ cross section  is larger than
the $(^1S_0)\;B_s$ cross section at all $p_T$,
however, their ratio is not precisely the naive prediction $3:1$,
but is about 20\% smaller.

In Fig.~\ref{fig2} we show the total cross section for the production of $B_s$
and $B_s^*$ mesons with transverse momentum above a minimum value
$p_T^{\rm min}$.  As in Fig.~\ref{fig1} only the range $p_T>5$ GeV and $|y|<1$
was considered.  The sensitivity to the scale $\mu$ was investigated, as
before,
by considering $\mu=\mu_R/2$ and $\mu=2\mu_R$, and the choice of this scale
is clearly not critical.  Figure \ref{fig2} provides useful estimates
of the production rates of $B_s$ and $B_s^*$ mesons at the
Tevatron due to the fragmentation process.  For convenience, we have also
presented these cross section estimates in Table~\ref{table}, including
the sum $\sigma(B_s)+\sigma(B_s^*)$.

The production rates and transverse momentum distributions for $B_s$ and
$B_s^*$ meson production in $p\bar p$ collisions presented here assume that
fragmentation is the dominant process and, therefore, are probably
underestimates.  At
large enough transverse momentum the fragmentation process should dominate,
as it falls off more slowly, even though it is only a part of the full order
$\alpha_s^4$ contribution.
We also note that in our calculation the light quark fragmentation
contribution is not included, since it is of even higher order in
$\alpha_s$ than the induced gluon fragmentation.  At the initial heavy
quark mass scale, the light quark fragmentation function $D_{q\to
B_s}(z,\mu)$ is of order $\alpha_s^4\, (\alpha_s^2)$ for
$q=u,d\,(q=s)$.  But the $D_{s\to B_s}(z,\mu)$ has been shown \cite{theory}
to be suppressed by $(m_s/m_b)^3$ relative to $D_{\bar b\to B_s}(z,\mu)$.
Therefore, at the initial scale none of these light quark
fragmentation functions are important.
The light quark fragmentation functions might acquire a logarithmic
enhancement from
the evolution of the Altarelli-Parisi equations, but the splitting kernels
$P_{g\to q}(x,\mu)$ and  $P_{q\to g}(x,\mu)$ are of order $\alpha_s$, while
$P_{b\to q}(x,\mu)$ and $P_{q \to b}(x,\mu)$ are of order $\alpha_s^2$.
Therefore, the most important sources of induced light quark fragmentation
come from
(i) $P_{q\to \bar b} \otimes D_{\bar b\to Bs}(z)$, and (ii)
    $P_{q \to g} \otimes D_{g \to B_s}(z)$.
Hence, the order of the induced light-quark fragmentation functions
at the scale $\mu$ can be at most of order $\alpha_s^4 \times \log(\mu/m_b)$,
which is suppressed by $\alpha_s$ relative to the induced gluon fragmentation.
Since the induced gluon fragmentation only contributes at about the 15--20\%
level, which is mostly due to the large gluon luminosity at low $x$,
the induced light-quark fragmentation contribution is expected to
be very small, at the level of a few \%.  And thus, the light quark
contribution can be safely ignored in compare with other uncertainties in the
calculation.

Our
fragmentation functions, which we used for the boundary conditions for the
Altarelli-Parisi equations at the scale of the heavy $b$ quark mass,
are rigorously correct in perturbative QCD if
the constituent quarks are much heavier than $\Lambda_{\rm QCD}$; e.g,
$B_c$ mesons.  Also higher order QCD corrections can, in principle,
be systematically calculated.  In the case of the $B_s$ meson,
higher order corrections in
the fragmentation functions might be large due to the fact that the
strong coupling constant is evaluated at $2m_s$ and is equal to about
0.75.  But we fitted the parameter $m_s$ to the experimental
measurement of the total probability for $\bar b \to B_s+B_s^*$ and $m_s$
turned out to be about 300 MeV$/c^2$.
We can, therefore, be confident of the overall normalization, which is
proportional to $\alpha_s^2(2m_s)|R(0)|^2/ m_s^3$.
The shapes of the fragmentation functions are only moderately sensitive
to the value of $m_s$ and comparing our results for the momentum distribution
will, therefore, primarily test the assumption that their {\em shapes} are
adequately given by perturbative QCD at the scale of the $b$ quark mass.

Our approach also appears somewhat similar to the approach used in
some Monte Carlo programs, e.g., PYTHIA \cite{pythia} and HERWIG \cite{herwig}.
However, the main difference lies in the fragmentation of the partons.
The fragmentation models used in PYTHIA or HERWIG are rather different
in spirit from our perturbative QCD fragmentation,
and is based on string fragmentation,
though the actual fragmentation models used in these two Monte Carlo
programs are somewhat different from each other.
In the string fragmentation picture, given
an initial $q_0$, it is assumed that a new $q_1 \bar q_1$ pair may be created
according to some pre-assigned probabilities such that a meson $q_0 \bar q_1$
is formed and a $q_1$ is left behind.  This $q_1$ may at a later stage pair
off with a $\bar q_2$, and so on.  The relative probabilities
used in PYTHIA to create $u\bar u,\, d\bar d,\, s\bar s,\, c\bar c$ pairs
are set at $1:1:\gamma_s:\gamma_c$, where $\gamma_s=0.3, \gamma_c=10^{-11}$
are default values.   This choice is based on a quantum mechanical effect and
is only approximate.
For example, in a paper by Lusignoli et al. \cite{plb},
in order to reproduce the $B_c$ meson production using HERWIG to within a
reasonable range, they had to increase the probabilities of creating $c\bar c$
pairs in the cluster splitting.  Thus, this ad hoc feature of string
fragmentation in these Monte Carlo programs
makes it less predictive than our perturbative QCD fragmentation
functions, because in our perturbative QCD fragmentation model
the probability for $\bar b \to B_s, B_s^*$ is
determined reliably once $m_b$, $m_s$, as well as the
nonperturbative parameter of the bound-state, e.g., the decay constant
$f_{B_s}$, are given.
The string fragmentation picture is rather phenomenological
since the parameter $\gamma_s$ can be adjusted freely to fit the experimental
measurements.
Another weakness of such a string fragmentation model is that
it must assume the $B_s/B_s^*$ ratio of $1/3$, from naive spin
counting, while our perturbative
QCD fragmentation model can reliably predict this
ratio once $m_s$ is given.  In fact, the ratio is slightly larger (20\%)
than $1/3$, as expected for a non-zero $m_s$, in the HQET \cite{pqcd} in
which $m_s/m_b$ is the leading heavy quark-spin symmetry breaking effect.

The comparison with forthcoming data from the
Fermilab Tevatron should be instructive.

\bigskip

This work was supported by the U.~S. Department of Energy, Division of
High Energy Physics, under Grant DE-FG02-91-ER40684 and DE-FG03-93ER40757.

\newpage

\begin{table}[h]
\caption[]{\label{table} \small
Total $B_s$ and $B_s^*$ cross sections in nb versus $p_T^{\rm min}$ for
$p_T>5$ GeV and $|y|<1$ at $\sqrt{s}=1.8$ TeV.  Variation
with the scale $\mu$ are shown.}
\medskip
\centering
\begin{tabular}{c|ccc|ccc|ccc}
$p_T^{\rm min}$  &  \multicolumn{3}{c|}{$\sigma(B_s)$} & \multicolumn{3}{c|}
{$\sigma(B_s^*)$} & \multicolumn{3}{c}{$\sigma(B_s)+\sigma(B_s^*)$} \\
\hline
  &  \underline{$\mu_R/2$} & \underline{$\mu_R$} & \underline{$2\mu_R$} &
     \underline{$\mu_R/2$} & \underline{$\mu_R$} & \underline{$2\mu_R$}  &
     \underline{$\mu_R/2$} & \underline{$\mu_R$} & \underline{$2\mu_R$}  \\
5   &   300 & 470 & 650 &    700 & 1100 & 1500  & 1000 & 1600 & 2200 \\
10  &    45 &  62 &  65 &    110 &  145 &  150  & 150  & 210  & 220 \\
15  &    14 &  16 &  14 &    32  &   37 &   33  &  46  &  52  &  47\\
20  &   5.2 &  5.1& 4.4 &    12  &   12 &   10  &  17.5&  17  &  15 \\
\end{tabular}
\end{table}

\newpage
\begin{center}
\section*{Figure Captions}
\end{center}

\begin{enumerate}

\item
\label{fig1}
The transverse momentum distributions of $B_s$ and $B_s^*$ mesons
for $p_T>5$ GeV, $|y|<1$, and $\mu=\mu_R/2, \mu_R$, and $2\mu_R$ at $\sqrt{s}
=1.8$ TeV.

\item
\label{fig2}
The $B_s$ and $B_s^*$ total cross sections for $p_T>p_T^{\rm min}$ with
$p_T>5$ GeV, $|y|<1$, and $\mu=\mu_R/2, \mu_R$, and $2\mu_R$ at $\sqrt{s}
=1.8$ TeV.

\end{enumerate}

\begin{figure}
\centering
\leavevmode
\epsfysize=430pt
\epsfbox{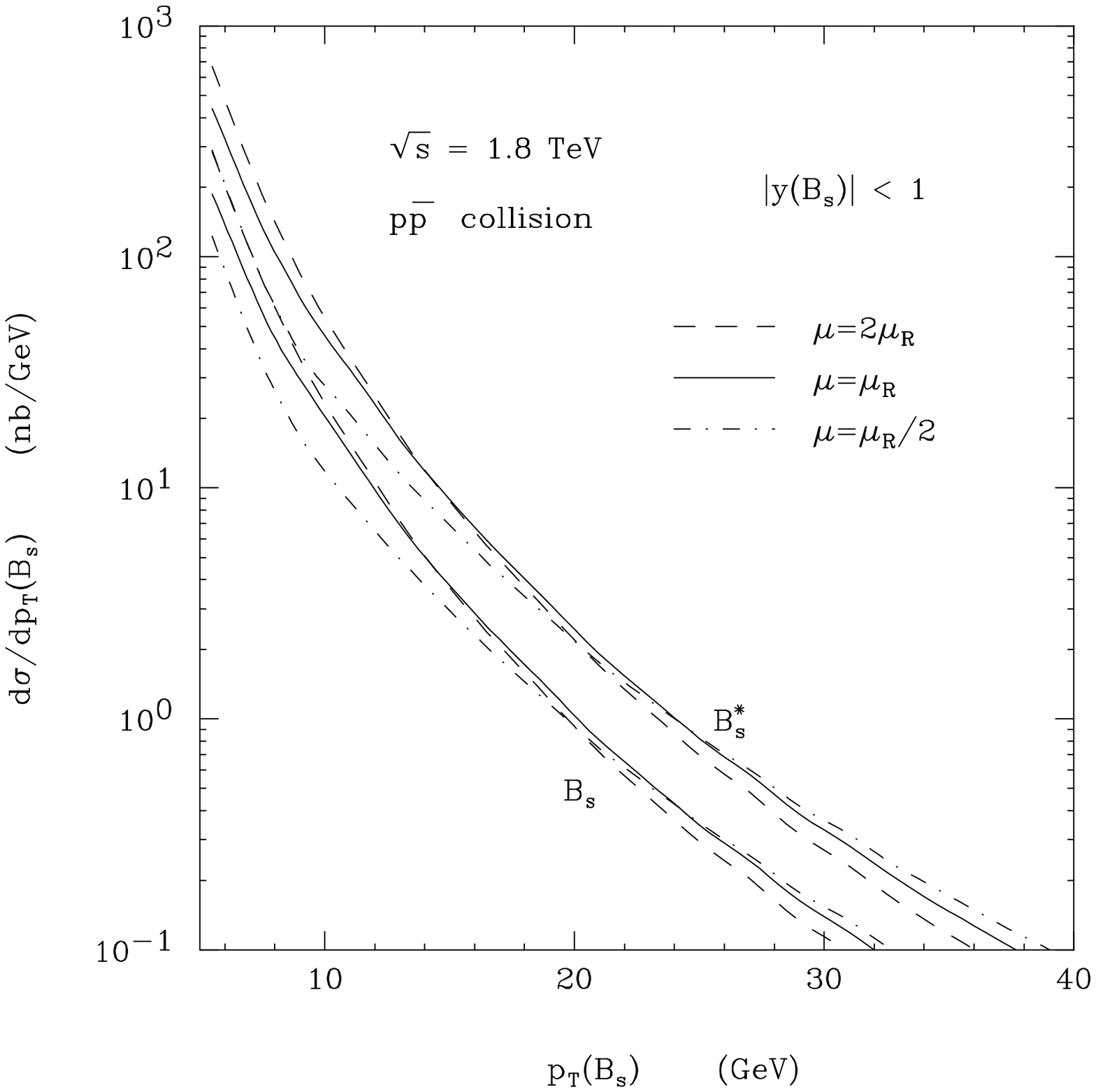}
\caption{}
\end{figure}

\begin{figure}
\centering
\leavevmode
\epsfysize=430pt
\epsfbox{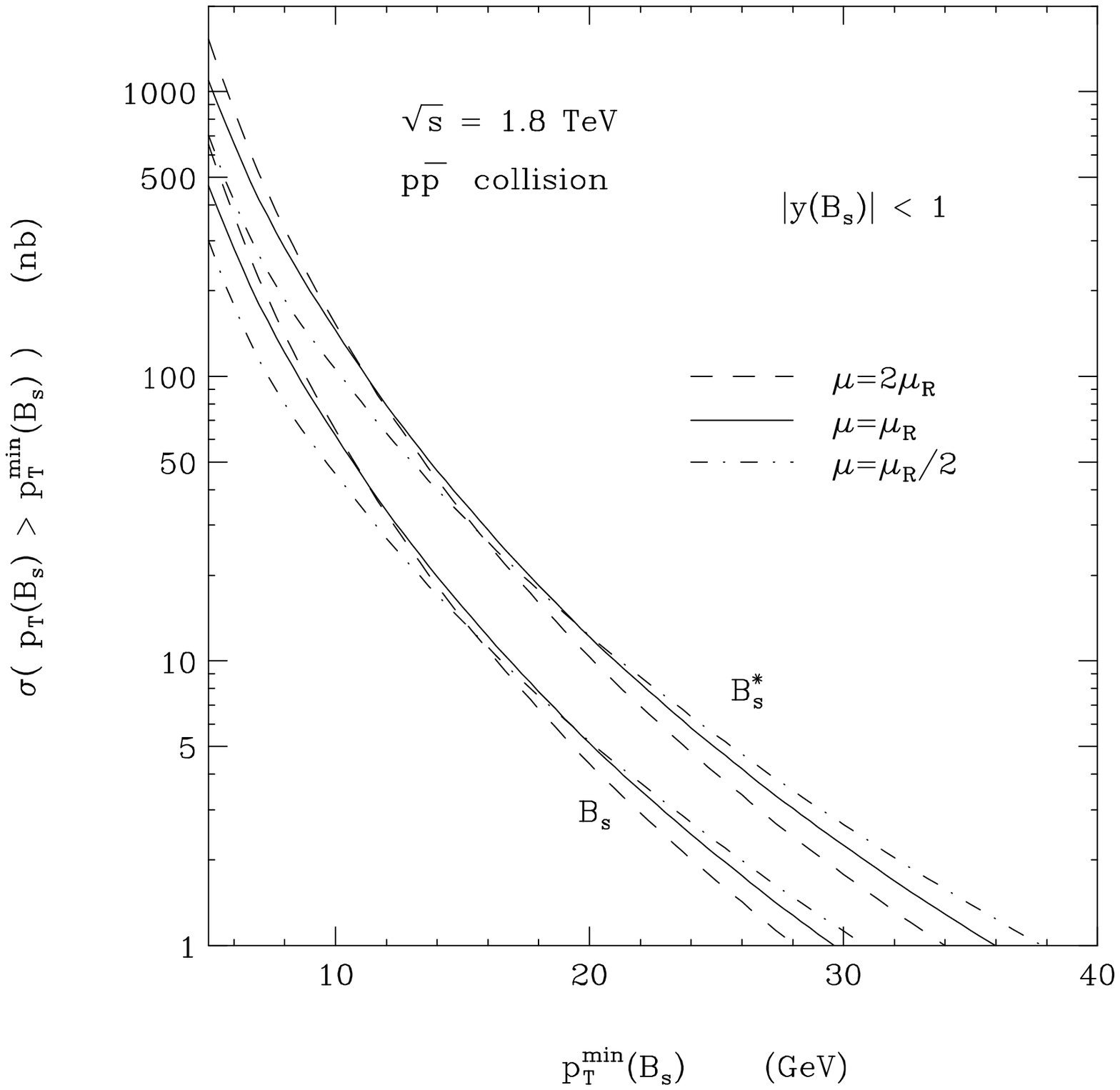}
\caption{}
\end{figure}


\newpage
\begin{thebibliography}{99}
%
\bibitem{delphi}P. Abrau {\it et al.} (DELPHI Collaboration),
Z. Phys {\bf C61}, 407 (1994).
\bibitem{bs}K. Cheung and R.J. Oakes, Phys. Lett. {\bf B337}, 181 (1994).
\bibitem{induce}K. Cheung, Phys. Rev. Lett. {\bf 71}, 3413,(1993);
K. Cheung and T.C. Yuan, Phys. Lett. {\bf B325}, 481 (1994);
K. Cheung and T.C. Yuan, preprint CPP-94-37 (Feb 1995), hep-ph/9502250.
%
\bibitem{cteq}H.L. Lai {\it et al.} (CTEQ Collaboration), Phys. Rev. {\bf D51},
4763 (1995).
\bibitem{pdg} Particle Data Group, Phys. Rev. {\bf D45}, 51 (1992).
\bibitem{alt} G. Altarelli and G. Parisi, Nucl. Phys. {\bf B126}, 298 (1977).
\bibitem{theory}E.~Braaten, K.~Cheung, and T.~C.~Yuan, Phys. Rev. {\bf D48},
R5049 (1993).
%
\bibitem{weiss}R. Van Royen and V. Weisskopf, Nuovo Cimento {\bf 50}, 617
(1967); {\bf 51}, 583 (1967).
%
\bibitem{soni}C. Bernard, J. Labrenz, and A. Soni, Phys. Rev. {\bf D49}, 2536
(1994).
%
\bibitem{old}DELPHI Coll., Phys. Lett. {\bf B289}, 199 (1992);
ALEPH Coll., Phys. Lett {\bf B294}, 145 (1992);
OPAL Coll., Phys. Lett. {\bf B295}, 357 (1992).
%
\bibitem{cdf}T.J. LeCompte (CDF Collaboration), submitted paper to the
27th Int. Conf. on High Energy Physics, Glasgow, Scotland, 20--27 July 1994,
FERMILAB-CONF-94/134-E.
%
\bibitem{pythia}{\it PYTHIA 5.7 and JETSET 7.4: Physics and Manual},
T. Sjostrand, hep-ph/9508391; Comput. Phys. Commun. {\bf 82}, 74 (1994).
%
\bibitem{herwig}G. Marchesini {\it et al.}, Comput. Phys. Commun. {\bf 67},
465 (1992).
%
\bibitem{plb}M. Lusignoli, M. Masetti, and S. Petrarca, Phys. Lett. {\bf
B266}, 142 (1991).
%
\bibitem{pqcd}E. Braaten, K. Cheung, S. Fleming, and T.C. Yuan, Phys. Rev.
{\bf D51}, 4819 (1995).
\end{thebibliography}
\end{document}